**Delegating Responsibilities to Intelligent Autonomous Systems: Challenges and Benefits**

Gordana Dodig-Crnkovic[1,2], Gianfranco Basti[3], and Tobias Holstein[1]

[1] Division of Computer Science and Software Engineering, Mälardalen University, Västerås, Sweden
[2] Department of Computer Science and Engineering, Chalmers University of Technology, Gothenburg, Sweden.
 e-mail: dodig@chalmers.se
[3] Pontifical Lateran University, Vatican City, Italy

**Abstract**

As AI systems increasingly operate with autonomy and adaptability, the traditional boundaries of moral responsibility in techno-social systems are being challenged. This paper explores the evolving discourse on the delegation of responsibilities to intelligent autonomous agents and the ethical implications of such practices. Synthesizing recent developments in AI ethics, including concepts of distributed responsibility and ethical AI by design, the paper proposes a functionalist perspective as a framework. This perspective views moral responsibility not as an individual trait but as a role within a socio-technical system, distributed among human and artificial agents. As an example of 'AI ethical by design,' we present Basti and Vitiello's implementation. They suggest that AI can act as artificial moral agents by learning ethical guidelines and using Deontic Higher-Order Logic to assess decisions ethically. Motivated by the possible speed and scale beyond human supervision and ethical implications, the paper argues for 'AI ethical by design', while acknowledging the distributed, shared, and dynamic nature of responsibility. This functionalist approach offers a practical framework for navigating the complexities of AI ethics in a rapidly evolving technological landscape.

**Introduction**

"(W)e need to ensure that we put in place the social and technical constructs that ensure responsibility and trust for the systems we develop and use in contexts that change and evolve. *Obviously, the AI applications are not responsible, it is the socio-technical system of which the applications are part of that must bear responsibility and ensure trust*. Ensuring ethically aligned AI systems requires more than designing systems whose results can be trusted. It is about the way we design them, why we design them, and who is involved in designing them. *This is a work always in progress*." (Dignum 2019) (emphasis added)

Even the United Nations Interim report, Governing AI for Humanity, published by the AI Advisory Body (UN 2023) emphasizes strongly the responsibility of humans for ethical AI, with the important first guiding principle: "*AI should be governed inclusively, by and for the benefit of all*".

From the systemic point of view, no doubt humans are the ones who are responsible for the building and maintenance of the whole socio-technological system and its ethics. However, in our previous work, (Dodig-Crnkovic and Çürüklü 2012) (Holstein, Dodig-Crnkovic, and Pelliccione 2018) (Holstein, Dodig-Crnkovic, and Pelliccione 2021a) (Dodig-Crnkovic et al. 2023) (Basti and Vitiello 2023) we elaborated on the necessity of building ethical assurance in the socio-technological system, *with distributed responsibilities, including delegating responsibilities to autonomous intelligent systems* (Dodig-Crnkovic and Persson 2008) (Candrian and Scherer 2022). It does not contradict the views of (Dignum 2019) and (UN 2023) but argues for the necessity of an intrinsic 'ethics by design' for autonomous intelligent artifacts. The reason for that is that autonomous intelligent artifacts learn and make decisions in ways their designers cannot predict. Instead of leaving them to follow their logic, priorities, and values, we should try to teach them and guide them in decision-making mechanisms so that their behavior aligns with human values. This article thus concentrates on the "machine ethics" part of the socio-technological system. That is, how ethics can be incorporated 'by design' in the artifact itself as part of a distributed agent system.

The analysis of distributed responsibility in such a network of agents as presented by (Taddeo and Floridi 2018) builds on the following argument pointed out by (Basti and Vitiello 2023):

"The effects of decisions or actions based on AI are often the result of countless interactions among many actors, including designers, developers, users, software, and hardware. This is known as a distributed agency. *With distributed agency comes distributed responsibility*.[1, p. 751]" (emphasis added)

Basti makes an important observation (Basti 2020) about the difference "between the *slow responsibility* of conscious ethical agents such as humans, and the *fast responsiveness* of unconscious skilled moral agents such as machines with respect to the ethical constraints from the shared social environment". (Basti and Vitiello 2023)

This difference in speed is indeed a factor we must take into account, along with the possibility of intelligent machines concurrently exchanging information on a global scale. In analogy with traditional safety-critical systems where redundancy is typically used to ensure safety, in the case of autonomous intelligent systems, given the speed beyond human supervision, one of the approaches together with 'ethicality by design,' may be adding redundancy, that is adding redundant AI systems as a way of control. Developing AI systems that can monitor other AIs in real time is one possible solution as shown by Basti and Vitiello. These monitoring systems could identify unusual or unethical behavior and alert human operators. Moreover, regular automated audits of AI behavior could be implemented to ensure compliance with ethical guidelines. These audits could be designed to run at a pace and scale that matches the speed of AI operations. In critical scenarios, distributing the decision-making process across multiple AI systems could ensure that no single system has the final say without input or verification from others. This could reduce the likelihood of unsafe decisions being made. Those questions remain to explore in the future, as AI systems gain increasing autonomy.

The key contribution of this work lies in its integration of current debates on AI responsibility with a novel synthesis of functional distributed responsibility and 'AI ethical by design'.

**Why do We Need to Embody Ethics in Intelligent Autonomous Systems?**

The European AI Act (EU 2024) defines an AI system as *"a machine-based system designed to operate with varying levels of autonomy and that may exhibit adaptiveness after deployment and that, for explicit or implicit objectives, infers, from the input it receives, how to generate outputs such as predictions, content, recommendations, or decisions that can influence physical or virtual environments",* which coincides the OECD's latest definition, (OECD.AI 2024). It emphasizes the role of autonomy, adaptiveness, learning, and decision-making.

Intelligent autonomous systems are becoming increasingly embedded in our daily lives, from personal assistants, smart home devices, healthcare robots, smart infrastructures, educational or vocational training systems, employment management applications, administrative tools, autonomous vehicles, scientific research, and publishing tools. They are part of industrial infrastructure, manufacturing and administration, agricultural robots, retail, and delivery, environmental monitoring, financial trading, entertainment, law enforcement, migration, asylum, and border control, digitalized democratic processes, surveillance systems, and military applications.

These systems do not operate in a vacuum; they make decisions that affect human lives, societal norms, and our perception of agency and responsibility. Consequently, given their ability to autonomously learn and make decisions, the urgency of embedding ethical principles in these systems grows.

Intelligent systems are designed to make decisions based on vast amounts of data and complex algorithms. Without ethical guidelines, these decisions can perpetuate biases, infringe on privacy rights, or lead to unequal treatment. The need for ethics is not just about preventing harm; it is about ensuring that technology augments the human experience without diminishing our dignity or autonomy and supports human flourishing.

However, even though 'AI ethical by design,' is important in ensuring the smooth delegation of sensitive tasks to AI there are constraints it must satisfy. For example, there are challenges in creating universally accepted ethical standards, as well as ethical dilemmas when different principles conflict, together with the difficulties in ensuring that these standards are adaptable to rapidly changing contexts. Diverse strategies can be used to alleviate those problems, such as dynamic ethical adjustment mechanisms that allow AI systems to update their ethical frameworks in response to new information or societal shifts.

**Can AI Be Morally Responsible? Functionalist Approach**

"Responsible Artificial Intelligence is about human responsibility for the development of intelligent systems along fundamental human principles and values, to ensure human flourishing and well-being in a sustainable world." (Dignum 2019)

Traditionally, moral responsibility is attributed to individuals based on their actions and intentions (Johnson and Powers 2005). But what about AI? Can machines with no real "minds" hold moral responsibility? Many argue that AI lacks the mental state of intention needed for accountability, and assigning praise or blame wouldn't have any meaning to them anyway. These views see responsibility as an individual trait. However, an increasing number of philosophers and scientists question the role of free will in moral responsibility.

"Whatever one makes of the relationship between free will and moral responsibility – e.g. whether it's the case that we can have the latter without the former and, if so, what conditions must be met; whatever one thinks about whether artificially intelligent agents might ever meet such conditions, one still faces the following questions. What is the value of moral responsibility? If we take moral responsibility to be a matter of being a fitting target of moral blame or praise, what are the goods attached to them? (…) I challenge this assumption by asking what the goods of this system, if any, are, and what happens to them in the face of artificially intelligent agents. I will argue that they neither introduce new problems for the moral responsibility system nor do they threaten what we really (ought to) care about." (Gogoshin 2024)

Along those lines, some philosophers, notably (Dennett 1973) propose a different, 'functionalist', approach. Dennett argues that responsibility is best seen as not just an individual property, but a social role shaped by group norms. *From this perspective, moral responsibility acts as a societal control system, encouraging good behavior and discouraging bad*.

The key point of Dennett's approach is that responsibility is a role that society assigns to individuals based on their function within a social structure. The social role of responsibility is shaped by normative expectations—what society expects of individuals in certain roles (e.g., lawmakers, teachers, engineers). These expectations are culturally dependent and evolving. This approach emphasizes the utility of holding someone responsible rather than focusing on metaphysical questions about free will or intentionality.

Adopting Dennett's 'functionalist' view Dodig-Crnkovic and Persson 2008) (Dodig-Crnkovic 2008) consider AI as part of larger, interconnected systems with shared/distributed responsibility. In this framework, responsibility exists on a spectrum, not as a binary yes or no variable.

In a "functionalist" perspective moral responsibility acts as a societal control system, emphasizing shared responsibility across the AI system and its stakeholders (Stahl 2023) (Lu et al. 2023) (de Laat 2021). By delegating tasks, (Candrian and Scherer 2022) (Hauptman et al. 2023) AI inherits responsibility for outcomes, making ethical considerations crucial for their intended function. This responsibility delegated to AI cannot focus on blame, but on encouraging good behavior and mitigating risk through constant learning in the socio-technological system, (Dodig-Crnkovic 2008).

**Responsibility as a Shared Social Role. Example of Ethics by Design for Autonomous Intelligent Robots**

With autonomous systems having ethical principles embedded into their decision-making processes, their role in society expands. They are not just tools or machines; they begin to take on an agential role. According to Dennett's functionalist approach, responsibility in this context is shared between the human stakeholders (designers, developers, users), and the intelligent autonomous systems themselves.

If autonomous systems are designed to follow ethical principles, they might be seen as holding a form of moral agency. From Dennett's perspective, this "agency" is not about the AI possessing free will, consciousness, or intentionality, but about fulfilling a socially assigned role that involves ethical decision-making.

The delegation of responsibility to these systems does not absolve humans of their role. Instead, it requires a new framework where responsibility is dynamically shared. Humans are responsible for designing, overseeing, and updating the ethical features programmed in these systems, while AI is responsible for the ethical execution of tasks.

In a future where autonomous systems operate with built-in ethics, the accountability structure becomes more complex. If an autonomous robot (including autonomous cars) with ethics by design causes an accident, determining responsibility would involve examining not just the human actors (producers, designers, users, regulators, etc.) but also the decisions made by the AI system within its ethical constraints.

Even with advanced ethical AI, human oversight of the techno-social system remains essential. Dennett's approach suggests that the ultimate responsibility still lies with humans who assign and monitor these social roles. The AI's role is to execute the tasks ethically, but the responsibility for the entire socio-technological system still falls on humans.

Since autonomous AI systems already started to be deployed, and their decisions affect society and individuals, it is not enough to have regulation and control only in the beginning, design phase, and the whole life cycle of technology. Designers, developers, and governing and controlling bodies define autonomous AI systems as they look at the beginning of their life cycle. But with time, the autonomous intelligent artifacts develop independently, learn, and change constantly so they make novel and unpredictable decisions. A socio-technological system is an ecosystem i.e. complex networks of interconnected agents, humans, machines, and organizations in constant mutual interactions, adaptations, and learning processes. Thus AI systems constituting technological ecologies within the socio-technological domain (Stix 2022)(Stix 2019), need to establish feedback loops and learn from experience.

*The socio-technological system must be built on continuous learning, of all parts– from the intelligent technology to the requirement specification, design, testing, etc.* (Holstein, Dodig-Crnkovic, and Pelliccione 2018) (Holstein, Dodig-Crnkovic, and Pelliccione 2021a) Being a part of the learning process of techno-social ecologies, the education of engineers and other stakeholders in professional ethics is fundamental for future development.

As Dignum says, "Obviously, errors will be made, and disasters will happen. More than assigning blame for these failures, we need to learn from them and try again, try better." (Dignum 2019)

*From a functionalist perspective, the focus shouldn't be on blame, but on ensuring good behavior in the future.* Moral responsibility as a 'regulation mechanism' can guide the development and use of AI for societal benefit. As (Floridi and Sanders 2004) noted, "We can avoid anthropocentric and anthropomorphic attitudes towards agenthood and rely on an *ethical outlook not necessarily based on punishment and reward but on moral agenthood, accountability, and censure*." (emphasis added)

The long-term, wide-ranging impacts of AI on society have a profound impact and require open democratic discussion, as these systems have the potential to fundamentally transform the future of humanity.

**Possibilities and Complexities of Sharing Responsibility with AI**

There are several strong arguments for building ethics into autonomous intelligent systems, making them "ethical by design" through machine ethics (Wallach and Allen 2009)(Anderson and Anderson 2011) (Winfield et al. 2019) (Thompson 2021)(Anderson and Anderson 2011)

The strengths of the machine ethics approach are formalization, scalability, mitigating risks, and avoiding harm. Embedding ethical principles helps make decisions that align with human values (such as trust and transparency, fairness and non-discrimination, and long-term sustainability), minimizing risks to individuals, society, and the environment. As technology becomes more integrated into our lives, protecting human values like privacy, autonomy, and freedom becomes crucial. Embedding ethics ensures that technology advances in a way that respects and upholds these core values.

However, building "ethical by design" systems is a complex challenge. Defining ethical principles, translating them into algorithms, and addressing potential conflicts between values are complex tasks. Furthermore, ethical frameworks need to be adaptable to different contexts and evolving societal norms. Reaching a consensus on ethical principles is difficult, and cultural and contextual differences exist. However, the example of autonomous cars shows that principles converge globally (Holstein et al. 2021). Moreover, there is no global consensus among humans on ethics, so the problem is not specific to AI.

Yet another challenge is accountability, i.e. the responsibility for the ethical decisions made by an autonomous system programmed with machine ethics. Legal frameworks and accountability mechanisms need to be developed.

The proposed complementary approaches include "Human-in-the-loop" systems where humans maintain ultimate decision-making authority, with AI assisting with well-defined tasks. The potential challenges are scalability, cost, and bias. Such systems might be particularly suitable for ensuring ethical AI in healthcare, autonomous vehicles, and finance. It is important to design humans-in-the-loop systems to ensure the human role is meaningful, engaging for humans, and not perpetuating existing biases, such as racial, gender, socioeconomic, data, and algorithmic biases.

Given the opacity of current AI systems and the need for humans to understand their decision-making, *explainable AI* is proposed for developing AI systems that can explain their decision-making processes, fostering trust and understanding in humans.

One challenge is the dynamic nature of both intelligent technology as well as ethics—what society considers ethical change over time. Autonomous systems would need to adapt to these changes, and the responsibility for ensuring they do so would rest with their human overseers. This could lead to continuous updates and recalibration of the ethical guidelines embedded within these systems.

As autonomous systems take on more ethical decision-making roles, the human role might shift from direct decision-makers to overseers and regulators of these systems. This could reduce the immediate burden of ethical decision-making on humans but increase the responsibility for ensuring these systems are properly designed, maintained, and updated.

Society might increasingly see AI systems as fulfilling certain roles traditionally held by humans, especially in situations requiring quick ethical judgments (e.g., emergency scenarios). Over time, this could lead to a new understanding of social roles and responsibilities, where AI systems are expected to behave ethically within the parameters set by humans.

In light of Dennett's approach, the future of autonomous cars and robots with built-in ethics by design suggests a shift towards a more distributed model of responsibility, where both human actors and AI systems share the social roles necessary for ethical decision-making. This approach aligns with the idea that responsibility is about fulfilling a function within a broader socio-technical system, rather than about individual free will or intent. While AI systems can be designed to follow ethical guidelines, the ultimate responsibility for these systems' behavior still rests with the humans who create, deploy, and regulate them. Thus, the role of humans will increasingly focus on the oversight and continual adaptation of these systems to ensure they remain aligned with societal values.

This perspective suggests that as AI and robotics continue to advance, our understanding of responsibility and ethics will need to evolve concurrently, embracing the complexities of shared responsibility in an increasingly automated and intelligentized world.

**Basti and Vitiello's Proposal for the Implementation of 'AI Ethical by Design'**

To illustrate a case of 'AI ethical by design,', we present the work of (Basti and Vitiello 2023) who discuss whether AI systems can be considered as 'artificial moral agents' that share responsibility with humans in ethical decisions. It begins by noting that human decisions, while ethically accountable, are made through brain processes that are not fully transparent to us. Similarly, even AI systems that are not fully transparent to us can be seen as ethically accountable if they meet two conditions:

- First, they must have *ethical guidelines built into their Machine Learning (ML) processes*.
- Second, *they must be able to assess their decisions ethically before acting*, using advanced reasoning capabilities.

This solution suggests using a specific type of logic to enable an ethical decision-making process in AI systems, found in both human brain processes and advanced AI networks. Effectively, this is an algebraic (Boolean) formalization of Kripke's relational modal logic i.e., of the logic of the different interpretations (semantics) of the necessity/possibility operators of the modal syntax. In this case, it is the *deontic* interpretation of the *obligation/permission*. Now, for satisfying the "Turing imitation game" in machine ethics, where a machine would behave in an ethical way aligned with human criteria, a two-step process is implemented like for humans, "before" and "after" the decision process, which is "opaque", both in humans and machines.

As a first step, we need the incorporation of "ethical constraints" into the supervised machine learning optimization algorithm. These constraints are ethical clauses to be satisfied in the optimization algorithm, to grant the first condition that AI decisions align with human moral values. In the case of AI systems for automatic trading in financial markets, the optimization process concerns the maximization of the profit. Then, the ethical clauses (ethical "and's" to be satisfied) could concern the origins and destinations of the capital to be invested.

As a second step, after this deontic First-Order Logic decision ("moral judgment") about the right individual action to be performed, the AI system needs a mechanism to assess ethically its decision before acting ("moral reasoning"). This is where a deontic Higher-Order Logic Ethical Reasoner comes into play (Benzmüller, Parenta, & van der Torre, 2020). It is an AI system capable of evaluating whether the decisions taken by another AI system endowed with ethical skills are compliant with a given set of ethical rules. Or, more generally, the Ethical Reasoner can perform an automatic and transparent ethical evaluation of the decisions (outputs) made by another symbolic or non-symbolic AI system, according to some set of ethical rules (e.g., the rules of the "AI Act" of the European Union) implemented in the Reasoner.

Indeed, the Ethical Reasoner is a symbolic AI system, fully "transparent" to human inquiry, able to give an open account of its ethical reasoning. Therefore, in the case that it is used as an ultimate layer (sub-system) of an AI system endowed with an opaque machine learning process with ethical constraints, the Ethical Reasoner could give an automatic transparent (positive or negative) assessment of the decisions performed by the system before they are transformed into actions. This could grant the transparent ethical accountability of the opaque decisions performed by an AI system with "ethical skills", which is necessary for defining it as "an artificial moral agent".

In short, this approach proposes teaching AI to make morally responsible decisions by first training it with ethical guidelines and then equipping it with a sophisticated logic system to evaluate its actions before it carries them out. This involves using advanced forms of logic and deep learning techniques that mimic human brain processes, enabling AI to assess the ethical implications of its decisions.

At this point, one may ask: what kind of ethical approach will AI take, given that different ethical theories (such as utilitarianism, deontology, virtue ethics, etc.) will lead to different strategies? Moreover, interpretations are often culturally based, and as has been argued, different cultures may have both different legal systems and different attitudes. This issue has been explored in a study on the ethics of autonomous cars, which demonstrates how ethical criteria for such vehicles tend to converge globally. (Holstein, Dodig-Crnkovic, and Pelliccione 2021b)

In the process of globalization of technology, the issues of harmonization have historically been approached through various methods, such as standardization and agreements. When communication across borders is necessary, stakeholders negotiate shared solutions. We can expect a similar approach to apply to globally diverse ethical value systems and cultural preferences.

**Conclusions**

The integration of intelligent autonomous systems into society presents profound ethical challenges, particularly regarding responsibility. By adopting a functionalist perspective, this paper argues that responsibility should be understood as a distributed role, shared among stakeholders within the socio-technical ecosystem. This perspective allows for a more nuanced understanding of how responsibility can be assigned and managed in systems where AI operates autonomously and at speeds beyond human control.

The Vitiello-Basti approach serves as an example in this context, demonstrating how AI systems can be designed with embedded ethical reasoning capabilities. By utilizing advanced logical frameworks, applied along with machine learning, this approach enables AI systems to assess their decisions within a moral context, aligning their actions with human values. Ethical AI by design is crucial in ensuring that autonomous intelligent systems operate within acceptable moral boundaries. However, the paper also emphasizes that ethical design alone is insufficient; it must be supported by continuous oversight, dynamic adaptation, and active societal involvement.

The functionalist perspective emphasizes the importance of shared responsibility, ethical design, and continuous learning as key components in the development of AI systems that are aligned with human values and capable of making decisions that support the well-being of society. This approach offers a path forward in the ethical integration of AI, ensuring that technology serves humanity.

Addressing moral responsibility in the age of AI requires a multifaceted approach spanning technical, legal, organizational, and societal domains. While significant work has been done in articulating ethical principles, more research is needed on effective implementation and governance mechanisms. These solutions need to be proactive, evolving alongside advancing AI capabilities to ensure a balance between leveraging AI's benefits and managing its potential risks.

This work contributes to the ongoing debates on AI responsibility, especially in light of the rapid development of increasingly powerful autonomous intelligent systems. It adopts a functionalist approach to distributed responsibility within a techno-social system, emphasizing the importance of designing intelligent agents that are "ethical by design" while leaving behind concepts like guilt and punishment as regulative tools, which have no relevance for machines.